\documentstyle[prl,aps,psfig]{revtex} 
\tighten
\draft
\begin{document}
\twocolumn[\hsize\textwidth\columnwidth
\hsize\csname@twocolumnfalse\endcsname

\preprint{\vbox{Submitted to Physical Review C}} 
\title{Relativistic Approach to Isoscalar 
       Giant Resonances in ${}^{208}$Pb}
\author{J. Piekarewicz}
\address{Department of Physics,
         Florida State University, 
         Tallahassee, FL 32306}
\date{\today}
\maketitle

\begin{abstract}
We calculate the longitudinal response of ${}^{208}$Pb using a
relativistic random-phase approximation to three different
parameterizations of the Walecka model with scalar
self-interactions. From a nonspectral calculation of the
response---that automatically includes the mixing between positive-
and negative-energy states---we extract the distribution of strength
for the isoscalar monopole, dipole, and high-energy octupole
resonances. We employ a consistent formalism that uses the same
interaction in the calculation of the ground state as in the
calculation of the response. As a result, the conservation of the
vector current is strictly maintained throughout the calculation.
Further, at small momentum transfers the spurious dipole
strength---associated with the uniform translation of the
center-of-mass---gets shifted to zero excitation energy and is cleanly
separated from the sole remaining physical fragment located at an
excitation energy of about 24~MeV; no additional dipole strength is
observed. The best description of the collective modes is obtained
using a ``soft'' parameterization having a compression modulus
of $K\!=\!224$~MeV. 
\end{abstract}
\vfill
]

\narrowtext

Almost forty years ago Thouless wrote a seminal paper on vibrational
states in nuclei in the random-phase approximation~\cite{Th61}. There
he showed how spurious states---such as those associated with a
uniform translation of the center-of-mass---separate out cleanly from
the physical modes by having their strength shifted to zero excitation
energy. Thirty years later Dawson and Furnstahl generalized Thouless'
result to the relativistic domain placing particular emphasis on the
role of consistency~\cite{DF90}. They showed how a fully
self-consistent approach guarantees the conservation of the vector
current as well as the decoupling of the spurious component of the
isoscalar dipole ($J^{\pi}\!=\!1^{-};T\!=\!0$) mode from the physical
spectrum. These results emerged after a careful treatment of the
negative-energy states. Indeed, neglecting their contribution resulted
in a violation of the vector current as well as the appearance of
substantial spurious strength in the response.  These fundamental
results emphasize that the Dirac single-particle basis is complete
only when positive- and negative-energy states are included.

Relativistic models of nuclear structure have evolved considerably
since they were first introduced by Walecka~\cite{Wa74} and later
extended by Serot~\cite{Se79}. Although the qualitative success of
these models relies almost exclusively on the dynamics generated by
the the scalar ($\sigma$) and vector ($\omega$) mesons, several
improvements have been introduced in order to enhance their
quantitative standing~\cite{SW86,SW97}. Chief among them is the
incorporation of scalar self-interactions which introduce important
non-linearities into the equations of motion. Perhaps the greatest
impact of these non-linear terms has been seen in the compression
modulus of nuclear matter. In a linear model the compression modulus
is predicted to be unreasonably large at $K\!=\!547$~MeV. Yet this
value can be reduced to $K\!=\!224$~MeV by the mere inclusion of
non-linear terms.  We will show here how this ``soft''
parameterization yields excitation energies for various compressional
modes in fair agreement with experiment.

While scalar self-interactions are now incorporated routinely into
most relativistic calculations of the nuclear ground state, their role
on the dynamics of the excited states is just being unraveled; most
calculations of the response of the mean-field ground state still use
the linear model. Applying non-linear models becomes technically more
difficult because the scalar-meson propagator no longer has a simple
Yukawa form. At present the only calculations of the response that
have incorporated non-linear terms are those by Ma and
collaborators~\cite{Ma97,Ma99}. One of the main conclusions of their
work is that {\it ``a large discrepancy remains between theory and
experiment in the case of the dipole compression mode''}.  We now show
that if one includes the full momentum dependence of the longitudinal
response, a unique physical fragment emerges at low momentum
transfer. This fragment---located at an excitation energy of
$E\!\approx\!24$~MeV---is identified as the isoscalar giant dipole
resonance (ISGDR).

We start from a Lagrangian having an isodoublet nucleon field ($\psi$) 
interacting via the exchange of isoscalar sigma ($\phi$) and omega 
($V^{\mu}$) mesons, an isovector rho ($b^{\mu}$) meson, and the 
photon ($A^{\mu}$). That is, the interacting Lagrangian density 
becomes~\cite{SW86,SW97}
\begin{eqnarray}
  {\cal L}_{\rm int} &=& 
    g_{\rm s}\overline{\psi}\psi\phi
   -g_{\rm v}\overline{\psi}\gamma^{\mu}\psi V_{\mu} 
     -  \frac{1}{2}g_{\rho}\overline{\psi}
        \gamma^{\mu}\tau_{a}\psi b^{a}_{\mu} \nonumber \\
    &-& \frac{1}{2}(1+\tau_{3})e\overline{\psi}
        \gamma^{\mu}\psi A_{\mu} - U(\phi) \;.
 \label{Lint}
\end{eqnarray}
In addition to meson-nucleon interactions the Lagrangian 
density includes scalar self-interactions of the form
\begin{equation}
 U(\phi) = \frac{1}{3!}\kappa\phi^{3}
         + \frac{1}{4!}\lambda\phi^{4} \;.
 \label{Uphi}
\end{equation}

Our theoretical program in the linear model has been described in 
great detail in several references~\cite{HP89,Pi90}. Here we merely
highlight the main features of the approach. The longitudinal response 
of the mean-field ground state is defined by
\begin{eqnarray}
  S_{\rm L}({\bf q},\omega) &=&
  \sum_{n}\Big|\langle\Psi_{n}|\hat{\rho}({\bf q})|
  \Psi_{0}\rangle\Big|^{2}
  \delta(\omega-\omega_{n}) \nonumber \\ &=&
  -\frac{1}{\pi}{\cal I}_m \Pi^{00}({\bf q},{\bf q},\omega) \;,
 \label{slong}
\end{eqnarray}
where $\hat{\rho}({\bf q})$ is the Fourier transform of the isoscalar
vector density, $\Psi_{0}$ is the exact nuclear ground state, and
$\Psi_{n}$ is an excited state with excitation energy
$\omega_{n}$. Note that the response is directly related to the
timelike polarization insertion $\Pi^{00}$. To compute the
linear response of the ground state of spherical nuclei---such as
${}^{208}$Pb---one starts by calculating ground-state properties in a
mean-field approximation. In this mean-field theory (MFT) nucleons
interact with the self-consistent field generated by all
positive-energy nucleons; vacuum loops are neglected in this
approximation. Such a calculation yields single-particle energies and
wave-functions for the occupied states as well as the mean-field
potential $\Sigma_{\rm MF}(x)$. It is precisely this mean-field
potential that one uses to compute the single-nucleon propagator
nonspectrally:
\def\lpmb#1{\mbox{\boldmath $#1$}}
\begin{equation}
 \Big(\omega\gamma^{0}\!+\!i{\lpmb\gamma}\cdot{\lpmb\nabla}
 \!-\!M\!-\!
 \Sigma_{\rm MF}(x)\Big)G_{\rm F}({\bf x},{\bf y}; \omega)\!=\!
 \delta({\bf x}-{\bf y}).
 \label{DiracEqn}
\end{equation}
There are several advantages in using a nonspectral representation 
for the nucleon propagator~\cite{HP89,Pi90,Sh89}. First, one avoids 
the artificial cutoffs and truncations that plague the spectral
approach~\cite{DF90}.  Second, both positive and negative-energy
continuua are treated exactly.  As a result, the contributions from
the negative-energy states to the response are included automatically.
Finally, a nonspectral evaluation of $G_{\rm F}$ poses no more
challenges, nor requires much more computational effort, than the
corresponding calculation of an individual single-particle
state. Having determined the occupied bound-state orbitals and the
nucleon propagator, the evaluation of the uncorrelated---or
single-particle---polarization becomes relatively 
straightforward~\cite{HP89,Pi90}.

To go beyond the simple single-particle response one must invoke the
relativistic random-phase approximation (RPA). The RPA builds
long-range coherence among the many particle-hole excitations with the
same quantum numbers by iterating the uncorrelated polarization to
infinite order~\cite{FW71}. Yet before going any further in the
description of the RPA we must stress two issues of paramount
importance. The first is consistency, which demands that the residual
particle-hole interaction used in the RPA be identical to the
interaction used to generate the mean-field ground state. Second, the
consistent relativistic response of the mean-field ground state
involves, in addition to the familiar particle-hole excitation, the
mixing of positive- and negative-energy states. These new
configurations are essential for the conservation of the vector
current and for the removal of spurious dipole strength from the
physical spectrum. Although in the MFT it is consistent to neglect
vacuum polarization~\cite{DF90}, the mixing between positive- and
negative-energy states remains of utmost importance.

The one new ingredient that we wish to add to our formalism is
scalar self-interactions. The added complication arises from the 
fact that the scalar-mediated interaction no longer has a 
simple Yukawa form. Rather, the scalar propagator now satisfies
a Klein-Gordon equation:
\begin{equation}
  \Big(\omega^{2}\!+\!\nabla^{2}\!-\!m_{\rm s}^{2}
  \!-\!U''(\phi)\Big)\Delta({\bf x},{\bf y};\omega)
  \!=\! \delta({\bf x}-{\bf y}) \;.
 \label{KGEqn}
\end{equation}
In infinite nuclear matter the scalar self-interactions introduce a
trivial modification: the scalar meson now propagates with an
effective mass $m_{\rm s}^{\star 2}\!=\!m_{\rm s}^{2}
\!+\!U''(\phi_{0})$, rather than with its free-space value. In the
finite system solving for the scalar propagator becomes technically
more difficult, but not more than solving for the 
nucleon propagator of Eq.~(\ref{DiracEqn}).  We have computed the
scalar propagator in momentum space and have expanded it in terms of
spherical harmonics so that the angular integrals appearing in the RPA
equations may be done analytically. A publication containing a more
detailed description of our techniques will be forthcoming.

\begin{figure}[h]
\leavevmode\centering\psfig{file=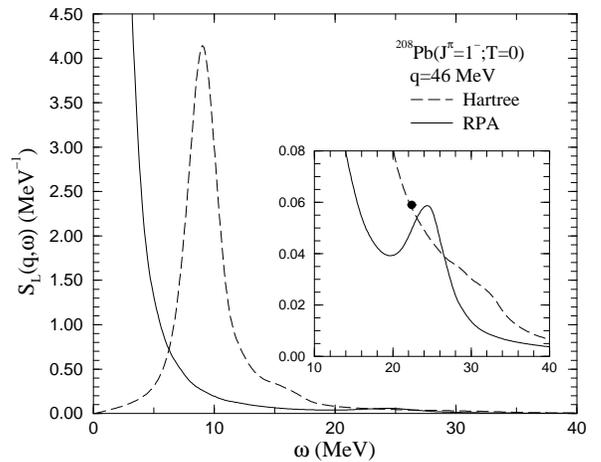,width=3in}
\caption{Distribution of isoscalar dipole strength in
         ${}^{208}$Pb at a momentum transfer of 
	 $q\!=\!46$~MeV. Calculations were done using 
	 parameter set NLC~\protect\cite{SW97} while
         the experimental value (filled circle on inset) 
         is from Ref.~\protect\cite{Da97}.}
 \label{Figure1}
\end{figure}

The benchmark by which every theoretical calculation of the nuclear
response should be measured is the isoscalar giant dipole resonance.
This is because the conservation of the vector current and the shift 
of spurious strength to zero excitation energy can only happen in a
consistent calculation of the response. In Fig.~\ref{Figure1} we
display the distribution of isoscalar dipole strength in ${}^{208}$Pb 
at the small momentum transfer of $q\!=\!46$~MeV 
(or $q\!=\!0.23~{\rm fm}^{-1}$) using parameter set NLC from 
Table~\ref{Table1}.
Note that the longitudinal response has been computed with an
``artificial'' width of 1~MeV. The uncorrelated Hartree response
displays a large amount of dipole strength around 8~MeV of excitation
energy. This strength is concentrated in the ``1-$\hbar\omega$''
region where many particle-hole excitations can be made. Yet most of
the strength is spurious, as revealed by the large amount being
shifted to zero excitation energy in the RPA response. What remains is
an almost imperceptible fragment located at $E=24.4$~MeV; and nothing
else. The small fragment is displayed more clearly along with the
experimental value (shown as a filled circle) on the inset of the 
figure. This result is a testimony to the power of consistency.  By
demanding that the residual particle-hole interaction be identical to
the interaction in the ground state, and by properly including the
mixing between positive- and negative-energy states, all spurious
strength gets cleanly separated from the physical response.

\begin{figure}[h]
\leavevmode\centering\psfig{file=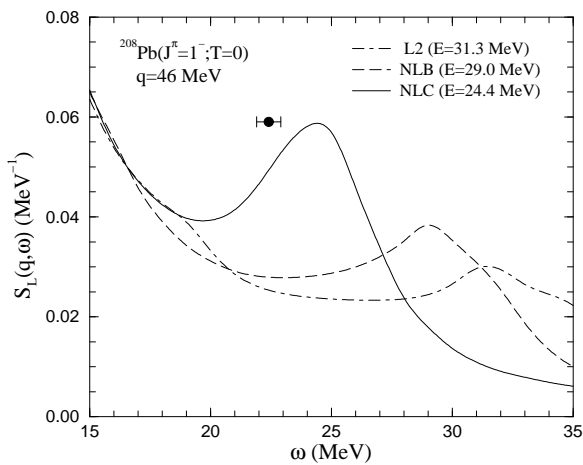,width=3in}
\caption{Distribution of isoscalar dipole strength in
         ${}^{208}$Pb at a momentum transfer of 
         $q\!=\!46$~MeV for three different models.
         The experimental value is from 
         Ref.~\protect\cite{Da97}.}
 \label{Figure2}
\end{figure}

\begin{figure}[h]
\leavevmode\centering\psfig{file=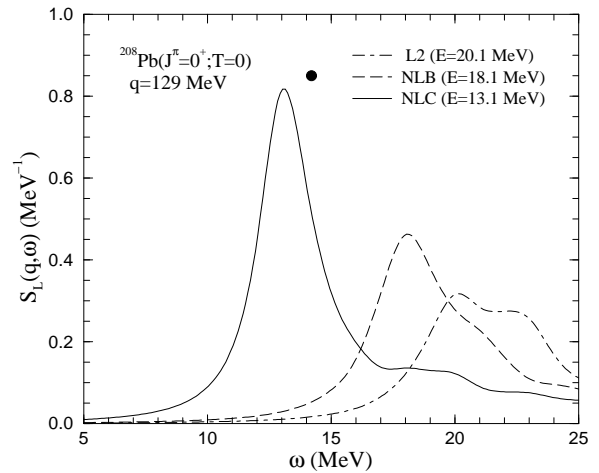,width=3in}
\caption{Distribution of isoscalar monopole strength in
         ${}^{208}$Pb at a momentum transfer of 
         $q\!=\!129$~MeV for three different models.
         The experimental value is from 
         Ref.~\protect\cite{Yo99}.}
 \label{Figure3}
\end{figure}

A comparison between three different relativistic models---all of them
constrained to reproduce bulk properties of nuclear matter at
saturation as well as the root-mean-square charge radius of
${}^{40}$Ca---is displayed in Fig.~\ref{Figure2}. Note that the three
models employed here have been defined in Ref.~\cite{SW97} as L2
($K\!=\!547$~MeV), NLB ($K\!=\!421$~MeV), and NLC ($K\!=\!224$~MeV)
[see also Table~\ref{Table1}]. As expected, the energy of the dipole
resonance scales with the compressibility of the model. Clearly,
models with a large compression modulus---such as L2 and NLB---produce
isoscalar dipole strength at values that are too large to be
consistent with experiment~\cite{Ad86,Da97}. These results have also
been tabulated in Table~\ref{Table2}. Because of the heroic efforts by
experimentalists in separating the isoscalar dipole mode from the
high-energy octupole resonance (HEOR), we also include a comparison
between our results and their experimental findings in
Table~\ref{Table2}. Although not necessarily a compressional mode,
our results for the HEOR follow similar trends to those observed 
for the giant dipole resonance.
We conclude the presentation of our results by displaying the
distribution of strength for the quintessential compressional mode:
the isoscalar giant monopole resonance (GMR). First discovered in
$\alpha$-scattering experiments from ${}^{208}$Pb~\cite{Yo77}, and
recently measured with higher accuracy at an excitation energy of
$E\!=\!14.2\pm0.1$~\cite{Yo99}, the GMR places important
constraints on theoretical models of nuclear matter. Indeed, the first
measurement of the GMR---in conjunction with a a simple analysis based
on the liquid-drop model---suggested a compression modulus of about
$K\!=\!200$~MeV, a value considerably lower than the predictions of
density-dependent Skyrme models at the time. Our calculations for the
monopole strength in ${}^{208}$Pb are displayed along with the
experimental value in Fig.~\ref{Figure3}. We find good agreement with
empirical formulas that suggest that the position of the GMR should
scale as the square root of the compressibility. Indeed, we compute 
GMR energies in the ratio of 1:1.38:1.53, while the square root of the
nuclear-matter compressibilities are in the ratio of 1:1.37:1.56. 
Moreover, these results help to reinforce our earlier claim that 
relativistic models of nuclear structure having compression moduli 
well above $K\!\approx\!200$~MeV will be in conflict with experiment.

In summary, we have computed the distribution of strength for the
isoscalar monopole, dipole, and high-energy octupole resonances in
${}^{208}$Pb using a relativistic random-phase approximation to three
different parameterizations of the Walecka model with scalar
self-interactions. We placed particular emphasis on the role of
consistency. That is, we demanded that the residual particle-hole
interaction used in the RPA be identical to the interaction used to
generate the mean-field ground state. Moreover, we have used a
nonspectral approach that automatically included the mixing between
positive- and negative-energy states to compute the longitudinal
response. Enforcing these constraints---and little else---was
sufficient for separating the spurious $J^{\pi}\!=\!1^{-};T\!=\!0$
state. In contrast to recent relativistic
calculations~\cite{Ma99}---as well as nonrelativistic
ones~\cite{VS81,Co99}---we see no need for imposing additional
constraints to ``partially'' remove the spurious contamination. These
approaches attempt to remove all spurious strength by defining an
effective dipole operator of the form $M_{10}({\bf
r})\!=\!(r^{3}\!-\!\eta r)Y_{10}(\hat{\bf r})$; here $\eta$ plays the
role of a Lagrange multiplier and is determined to be
$\eta\!=\!5\langle r^{2}\rangle/3$ from translational invariance. More
significantly, such a transition operator neglects the all important
momentum dependence of the excitation. Indeed, it was only at small
momentum transfers---just as in the experiment~\cite{Da97}---that we
observed a single physical fragment concentrated around 24 MeV of
excitation. As the momentum transfer increased, we uncovered
additional dipole strength around 8 MeV. Yet this trend, namely, a
sizable fraction of dipole strength at low energies and a giant
resonance peak, is all that was reported in those recent
publications~\cite{Ma99,Co99}. We do not regard this behavior as {\it
``a large discrepancy between theory and experiment''}. Rather, we
attribute this trend to the
simplified---momentum-independent---choice of dipole operator
adopted in those calculations.

We have also computed the distribution of strength for the giant
monopole resonance. As in the case of ISGDR we have used the same
exact operator---the isoscalar vector density---to compute the
monopole component of the longitudinal response. Indeed, monopole,
dipole, and octupole strength were all obtained from simply isolating
the relevant $J^{\pi}$-channel from the longitudinal response. For 
the GMR we have found good agreement with a recent relativistic
calculation~\cite{Ma97}. Good agreement has also been obtained with
semi-empirical formulas that suggest that the position of the GMR
should scale as the square root of the compressibility.  Depending on
the relativistic parameterization adopted, monopole strength was found
between 13 and 20 MeV of excitation.

Lastly, we venture into the neutron-star domain. The non-linear
parameterization NLC gives a rather satisfactory description of the
various compressional modes. Although by no means perfect, this
agreement suggests that the compression modulus of nuclear matter can
not differ too much from the value predicted by this model
($K\!=\!224$~MeV). When this
parameter set is used to compute the equation of state for neutron
matter---and is then combined with the Tolman-Oppenheimer-Volkoff
equation---one obtains an upper limit for the mass of a neutron star
of $M\!=\!2.8M_{\odot}$.  Given the recent compilation of 21
neutron-star masses by Thorsett and Chakrabarty~\cite{TC99}, where
they show that the measurements are consistent with a remarkably
narrow mass distribution $M=(1.35\pm 0.04) M_{\odot}$~\cite{TC99}, the
fascinating possibility that neutron stars harbor novel and exotic
states of matter becomes almost a reality.

\medskip
This work was supported in part by the DOE under Contract 
No.DE-FG05-92ER40750 and by the Florida State University
School of Computational Science and Information Technology. 

\begin{table}
\caption{Various relativistic parameter 
         sets~\protect\cite{SW97}. The
         scalar mass and $\kappa$ are in 
	 MeV.} 
 \begin{tabular}{crcccrr}
 Set & $g_{\rm s}^{2}$ & $g_{\rm v}^{2}$ 
     & $g_{\rho}^{2}$  & $m_{\rm s}$ 
     & $\kappa$ & $\lambda$ \\
 \hline 
 L2    & 109.63 & 190.43 & 65.23 & 520 &    0 &    0 \\
 NLB   &  94.01 & 158.48 & 73.00 & 510 &  800 &   10 \\
 NLC   &  95.11 & 148.93 & 74.99 & 501 & 5000 & -200 
  \label{Table1}
 \end{tabular}
\end{table}
\vskip-0.2in
\begin{table}
\caption{Energies for various isoscalar resonances 
         in three different relativistic models. 
         All excitation energies are given in MeV.}
 \begin{tabular}{cccc}
 Model & GMR   & ISGDR  &  HEOR     \\
 \hline 
 L2    & 20.1  & 31.3   &  25.1     \\
 NLB   & 18.1  & 29.0   &  23.4     \\  
 NLC   & 13.1  & 24.4   &  21.9     \\ 
 \hline
 Exp.  & 14.2 $\pm$ 0.1\protect\cite{Yo99}
       & 22.4 $\pm$ 0.5\protect\cite{Da97} 
       & 19.7 $\pm$ 0.5\protect\cite{Da97}
  \label{Table2}
 \end{tabular}
\end{table}
\vskip-0.3in


\begin{references}
\bibitem{Th61}  D.J Thouless, Nucl. Phys. {\bf 22}, 78 (1961).
\bibitem{DF90}  J.F. Dawson and R.J. Furnstahl,
                Phys. Rev. {\bf C42}, 2009 (1990).
\bibitem{Wa74}  J.D.~Walecka, 
	        Ann. of Phys. {\bf 83}, 491 (1974).
\bibitem{Se79}  B.D. Serot, Phys. Lett. {\bf 86B}, 146 (1979).
\bibitem{SW86}  B.D. Serot and J.D. Walecka,
	        Adv. in Nucl. Phys. {\bf 16}, 
	        J.W. Negele and E. Vogt, eds. 
                (Plenum, N.Y. 1986).
\bibitem{SW97}  B.D. Serot and J.D. Walecka,
                Int. Jour. Mod. Phys. 
                {\bf E6}, 515 (1997).
\bibitem{Ma97}  Zhongyu Ma, Nguyen Van Gai, Hiroshi Toki, 
                and Marcelle L'Huillier,
                Phys. Rev. {\bf C55}, 2385 (1997).
\bibitem{Ma99}  Zhongyu Ma and Nguyen Van Gai, 
                {\tt nucl-th/9910054}.
\bibitem{HP89}  C.J. Horowitz and J. Piekarewicz,
                Phys. Rev. Lett. {\bf 62}, 391 (1989);
		Nucl. Phys. {\bf A511}, 461 (1990). 
\bibitem{Pi90}  J. Piekarewicz,
		Nucl. Phys. {\bf A511}, 487 (1990). 
\bibitem{FW71}  A.L. Fetter and J.D. Walecka,
	        {\it ``Quantum Theory of Many Particle
	        Systems''} (McGraw-Hill, New York, 1971).
\bibitem{Sh89}  J.R. Shepard, E. Rost, and J.A. McNeil,
	        Phys. Rev. {\bf C40}, 2320 (1989).
\bibitem{Ad86}  G.S. Adams, T.A. Carey, J.B. McClelland,
                J.M. Moss, S.J. Seestrom-Morris, and
	        D.Cook, Phys. Rev. {\bf C33}, 2054 (1986).
\bibitem{Da97}  B.F. Davis {\it et al.,}
		Phys. Rev. Lett. {\bf 79}, 609 (1997).
\bibitem{Yo77}  D.H. Youngblood, C.M. Rozsa, J.M. Moss,
                D.R. Brown, and J.D. Bronson,
                Phys. Rev. Lett. {\bf 39}, 1188 (1977).
\bibitem{Yo99}  D.H. Youngblood, H.L. Clark, and Y.-W. Lui, 
                Phys. Rev. Lett. {\bf 82}, 691 (1999).
\bibitem{VS81}  Nguyen Van Gai and H. Sagawa,
		Nucl. Phys. {\bf A371}, 1 (1981). 
\bibitem{Co99}  G. Col\`o, N. Van Gai, P.F. Bortignon,
                and M.R. Quaglia, {\tt nucl-th/9904051}.
\bibitem{TC99}  S.E. Thorsett and Deepto Chakrabarty,
                Ast. Jour. {\bf 512}, 288 (1999).

\end{references}
\end{document}